\documentclass[3p,times]{elsarticle}

\usepackage{ecrc}
\usepackage{caption}
\usepackage{subcaption}

\volume{00}

\firstpage{1}

\journalname{Nuclear Physics A}

\runauth{A.D\'{a}vila}

\jid{nupha}

\usepackage{amssymb}

\usepackage[figuresright]{rotating}

\begin{document}

\begin{frontmatter}



\dochead{}

\title{Proton/pion ratios in $\Delta\phi$ with respect to a jet in $\sqrt{s_{NN}}$=200 GeV Au+Au collisions at STAR  }


\author{A. D\'{a}vila for the STAR Collaboration}
\ead{alandav@physics.utexas.edu}

\address{The University of Texas at Austin,Physics Department, 1 University Station C1600,Austin, Texas 78712-0264}

\begin{abstract}
Inclusive proton/pion ratios show an enhancement at intermediate transverse momentum ($p_{T} \sim 1.5$ - 4.0 GeV/c) in central $\sqrt{s_{NN}}$ = 200 GeV Au+Au collisions compared to peripheral Au+Au and p+p collisions. This effect suggests a production mechanism, different from fragmentation, which is consistent with coalescence and recombination models. A high $E_{T}$ trigger particle selects a surface-biased jet, which is measured to have a similar $p_{T}$ distribution as a p+p jet. This jet is used to enhance the quenching effects of the recoiling, medium traversing one. We reconstruct the trigger jet using the FASTJET algorithm, with a $E_{T}$ ($p_{T}$) cut of 3.0 GeV(/c) on the towers(tracks) in order to reduce the heavy-ion collision background. The particle identification of tracks with $p_{T}$ up to $\sim$ 2.8 GeV/c is obtained by taking advantage of STAR TOF and TPC detectors with full azimuth coverage. Correlations in $\Delta\phi$ between jets and identified hadrons are presented, and the particle ratios in different regions of azimuth are measured. Particle ratios associated with the trigger jet vs. the recoil jet, and comparisons to inclusive particle ratios can help to distinguish between jet-related (vacuum and medium-modified) and bulk-related contributions to the ratios enhancement.

\end{abstract}


\end{frontmatter}


\section{Introduction}
\label{1}
The particles' production mechanism in heavy ion collisions has been tested with measurements of $p_{T}$ dependent particle ratios. The enhancement  of p/$\pi$ (\={p}/$\pi^{-}$) at intermediate $p_{T}$ and mid-rapidity was first observed in central $\sqrt{s_{NN}}$ = 130 GeV Au+Au collisions where the p (\={p}) yield becomes comparable to the $\pi$ ($\pi^{-}$) yield at $p_{T}$ $\sim$ 1.6 (2.0) GeV/c \cite{PhysRevLett.88.242301}. This enhancement had not been observed in peripheral and p+p collisions at a wide range of energies. The discovery was followed by further investigations by the PHENIX and STAR collaborations. The enhancement persists at $\sqrt{s_{NN}}$ = 200 GeV. Measurements at PHENIX showed that the p/$\pi$ ratio increases with $p_{T}$ up to $\sim$ 3.0 GeV/c in the 0-10 \% most central Au+Au collisions while it increases and saturates with a value $\sim$ 0.4 at $p_{T}\sim 1.5$ GeV/c in the 60 - 92 \% centrality \cite{PhysRevC.69.034909}. The ratio of central over peripheral binary scaled $p_{T}$ spectra ($R_{CP}$) of protons is close to 1 in the $p_{T}$ range of $\sim$ 1.5 - 4.0 GeV/c. This suggests that protons are either less suppressed than pions or that the production mechanism in central Au+Au collisions leads to more copiously produced protons than expected from fragmentation. It was argued then that a good explanation of particle production in this $p_{T}$ region would require more than just hydrodynamics or pQCD (fragmentation) \cite{PhysRevC.69.034909}.    
\\
\indent The p/$\pi$ measurements were extended to a higher $p_{T}$ by the STAR collaboration, reaching the region where particle production is dominated by jet fragmentation ($p_{T} \geq$ 6 GeV/c) \cite{PhysRevLett.97.152301}. The new measurements showed the complete $p_{T}$ neighborhood where the unexpectedly high p/$\pi$ is found. The p/$\pi$ ratio gets enhanced at intermediate $p_{T}$ (1.5 - 4.5 GeV/c) in 0-12 \% most central Au+Au collisions compared to peripheral (60-80\%) and d+Au collisions at $\sqrt{s_{NN}}$= 200 GeV. The ratio peaks at $\sim 2-3 $ GeV/c in central Au+Au and then decreases to approach the value measured in peripheral and d+Au collisions at about $p_{T}$ = 5 GeV/c. The fact that the p/$\pi$ ratio is the same at this high $p_{T}$ suggests that the hadron production mechanism is the same in central and peripheral collisions and that the partons fragmenting into the final state proton and pion lose the same amount of energy while passing through the medium in central Au+Au collisions.
\\
\indent There have been attempts to explain the p/$\pi$ anomalous enhancement by coalescence or recombination models and partonic energy loss effects on the p and $\pi$ spectrum \cite{PhysRevLett.90.202302}\cite{PhysRevLett.90.202303}\cite{TrainorIJMPE08}. It is argued \cite{PhysRevLett.90.202303} that the recombination of partons is the predominant hadron production mechanism whenever they follow a thermal distribution. Fragmentation becomes the dominant mechanism once the parton distribution becomes a power law. The recombination model correctly predicted a decrease of p/$\pi$ ratio at $p_{T} \sim$ 4-5 GeV/c and saturation afterwards once the fragmentation processes take over \cite{PhysRevC.68.044902}. A very similar coalescence model \cite{PhysRevLett.90.202302} proposes coalescence of minijet partons with thermal quarks as the source of the p/$\pi$ enhancement. The \={p}/$\pi^{-}$ ratio in central $\sqrt{s_{NN}}$ = 200 GeV Au+Au collisions is well described up to $p_{T}$ = 4.0 GeV/c with a suitable choice inverse slope parameter for the intermediate thermal antiprotons. The same coalescence model predicted an increase (decrease) of p/$\pi$ (\={p}/$\pi^{-}$) at intermediate $p_{T}$ when going from 200 to 62 GeV \cite{PhysRevC.71.041901}. STAR measurements confirmed the general predictions of these two models but showed some quantitative disagreement between models and data \cite{Abelev2007104}. Another suggested explanation of the p/$\pi$ enhancement places its origin on the effect of parton energy loss on the hard component of particle spectrum of Au+Au collisions. The measured spectra are described by a soft (scaling with number of participant pairs) + a hard (scaling with number of binary collisions) components. The hard part is isolated and its modification with respect to $N_{bin}$ scaling is related to a negative boost in the hard component hadrons' $p_{T}$. The modification of the hard component shows up as an increase of protons in the $p_{T}$ region of p/$\pi$ enhancement \cite{TrainorIJMPE08}
\\
\indent We propose to improve our understanding of the production mechanism giving rise to the p/$\pi$ enhancement by studying this ratio in a vacuum fragmenting jet and a medium traversing one. This will allow us to look for medium modification effects on the p/$\pi$ ratio. We will study 0-20 \% most central Au+Au collisions at $\sqrt{s_{NN}} = 200$ GeV and compare the particle production in the direction of a surface bias (near side) jet and its recoiling partner (away side) in the $p_{T}$ region where the p/$\pi$ ratio peaks. 

\section{STAR detector}
\label{2}
The STAR detector is based at the Relativistic Heavy Ion Collider in Brookhaven National Laboratory. It has charged and neutral particle detection capabilities with full azimuth and pseudorapidity coverage of approximately +- 1.5 units around mid-rapidity. The charged particles' momenta are reconstructed using the Time Projection Chamber (TPC). The TPC measures the P10 gas (10\% methane, 90\% argon) ionization energy loss by particles coming out of the collision. The energy lost is compared to the Bichsel expectation value for particle identification (PID) purposes \cite{Anderson2003659}. Gammas and $\pi^{0}$'s are detected via the Barrel Electromagnetic Calorimeter (BEMC) which is composed of 4800 towers of size 0.05 x 0.05 in $\Delta\eta, \Delta\phi$. Each tower is made of alternating lead and scintillator material that is used to measure the transverse energy deposition through radiation \cite{Beddo2003725}. Any track pointing toward a tower can induce radiation energy; therefore, the total momentum of matching tracks (or total energy if the track corresponds to an electron) is subtracted to avoid double counting. The Time of Flight detector had complete azimuth coverage starting on the Run 10 of STAR \cite{Shao2006419}. Along with the TPC it allows better PID in the region of overlap of the dE/dx expected values for kaon, protons and pions (0.7 -4.0 GeV/c). 
\\
\indent The data used for this analysis consisted of 9.4 million high tower ($E_{T} >$  4.3 GeV) triggered events with a 0-20 \% centrality cut from the Run 10 $\sqrt{s_{NN}}$ = 200 GeV Au+Au. An extra offline cut was imposed by requiring a high momentum jet with a neutral particle of $E_{T} = 5.0$ GeV within it reducing the statistics to 167K events. A collision vertex cut of 30.0 cm in the longitudinal direction was also imposed to get uniform acceptance. Pile up was reduced by rejecting events with an abnormally high number of global tracks relative to the number of primary tracks. 

\section{Jet trigger and particle identification }
\label{3}
We use a high $E_{T}$ particle to select a surface biased jet. In particular, we are interested in the $\Delta\phi$ correlation of PID particles with respect to a jet axis. We therefore isolate the azimuth direction of a parton coming from a hard scattering by making use of the FASTJET jet reconstruction algorithms \cite{Cacciari:2011ma}\cite{Cacciari:2005hq}. We reconstruct jets with the anti-kt jet finding algorithm and resolution parameter 0.4. A tower (track) cut of 3.0 GeV(/c) is imposed on all the particles used for jet reconstruction to reduce the heavy ion background. A $E_{T}$ = 5 GeV tower is required to be present in the jet. Only jets with $8 < p_{T}^{jet} \leq$ 20 GeV/c are accepted for further analysis. The cuts imposed on the jet trigger bias the jet population towards non interacting jets. The recoiling jet is therefore biased to interact with the medium. Note that the trigger $p_{T}^{jet}$ is not necessarily equal to the parton $p_{T}$. We study the intermediate $p_{T}$ component of the jet via the $\Delta\phi$ correlations at a later stage. The $p_{T}$ distribution of the triggered jets is measured to be similar to jets from high tower triggered p+p events at the same energy \cite{AliceQM2011}.
\\
\indent Protons and pions are identified probabilistically up to $p_{T}$ = 2.8 GeV/c using information from the TPC and TOF detectors. The number of sigmas ($n\sigma$) is defined as the difference of the log (dE/dx) given a particle mass assumption and the Bichsel expected value divided by the dE/dx resolution. Hadron's velocities can be obtained using the TOF information. The $\Delta\beta^{-1} / \beta^{-1} $ variable is defined as $\Delta\beta^{-1} / \beta^{-1} = (\beta^{-1}_{TOF}-\beta^{-1}_{TPC})/(\beta^{-1}_{TOF})$ where $\beta^{-1}_{TOF}$ is the inverse of the particle's velocity measured by TOF and $\beta^{-1}_{TPC}$ is calculated from the TPC measured momentum and using a pion(kaon or proton) mass assumption. Combining the $n\sigma$ with $\Delta\beta^{-1} / \beta^{-1}$ we can obtain 2D distributions where kaons, protons and pions can be distinguished. The 2D distributions are fit with a 2D model consisting of a Gaussian in the $n\sigma$ variable and a Student-t distribution in the $\Delta\beta^{-1} / \beta^{-1}$ variable for each particle (kaon, proton and pions). The fit is done in different $\eta$ slices of width $\delta\eta$ = 0.2 and momentum slices of $\Delta p$ = 0.2 GeV/c. Each $\eta$ and p distribution has its model parameters which are used to construct a PID probability function. The probability function is obtained by dividing the model function of the particle of interest by the sum of the model functions of all the particles present in the fit. The parameters of the model fits are obtained by using the particles distributions of 9.2 M events from the high tower $\sqrt{s_{NN}}$ = 200 GeV Au+Au data with a centrality of 0-20 $\%$. 
\begin{figure}[ht]
\begin{minipage}[b]{0.5\linewidth}
\centering
\includegraphics[width=\textwidth]{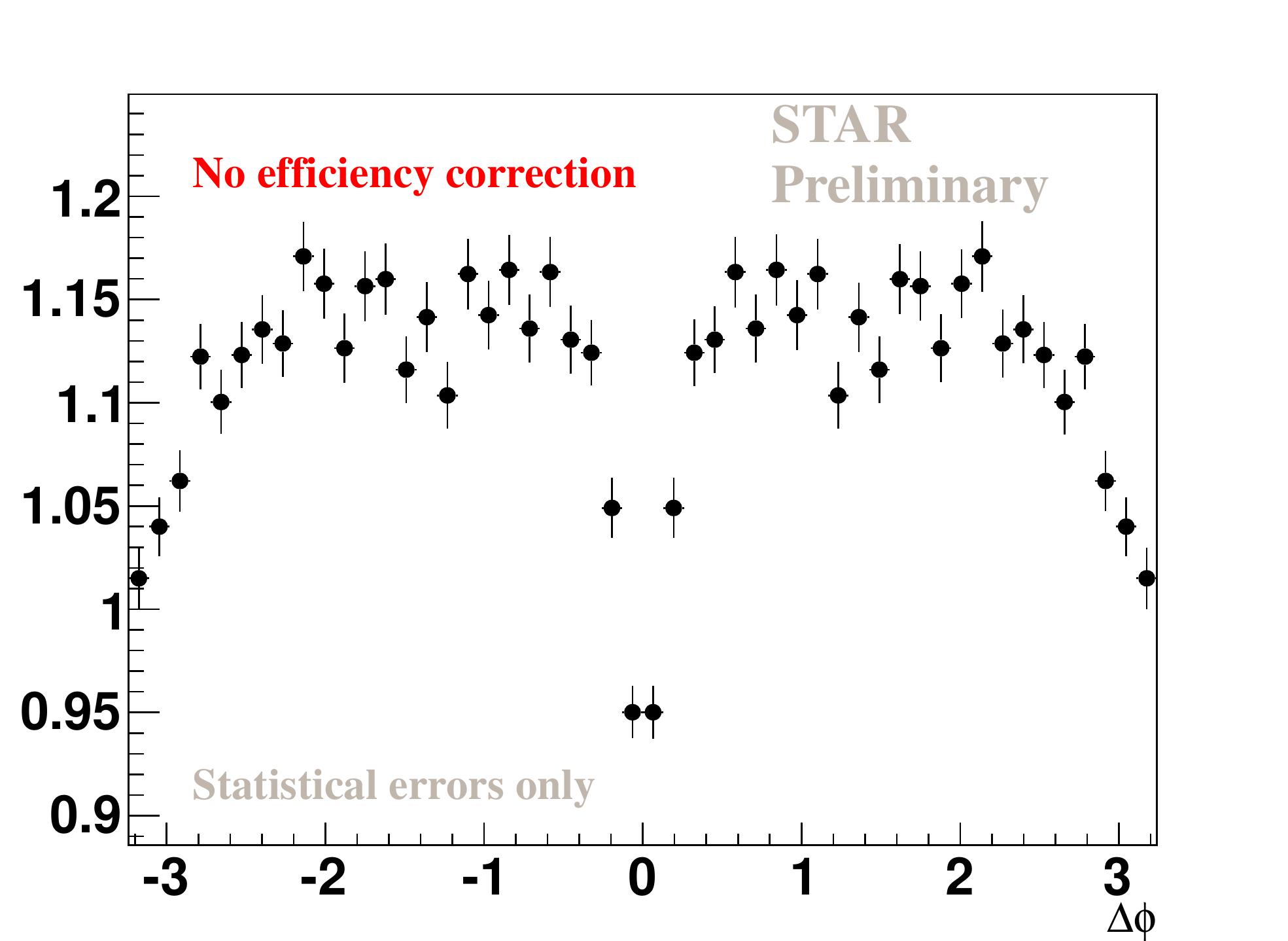}
\captionof{figure}{p/ $\pi$ ratio at 2.0 $\leq p_{T} < 2.8 $ GeV/c as function of $\Delta\phi$ with respect to a $8 < p_{T}^{jet} \leq 20 $ GeV/c trigger jet axis in $\sqrt{s_{NN}} = 200$ GeV 20 $\%$ most central Au+Au collisions.}
\label{fig:simpleRatio}  
\end{minipage}
\hspace{0.5cm}
\begin{minipage}[b]{0.5\linewidth}
\centering
\includegraphics[width=1.0\linewidth]{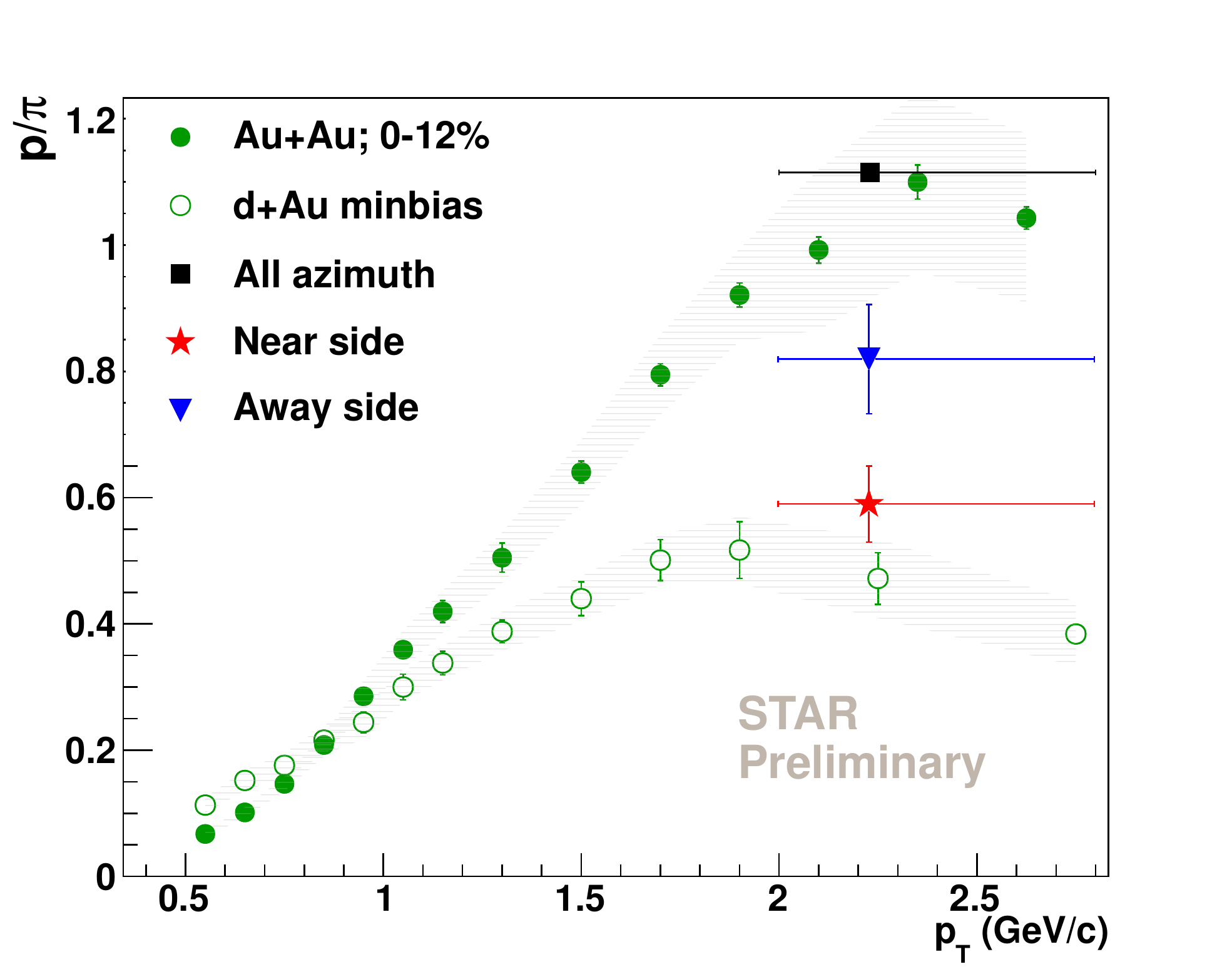}
\captionof{figure}{p/$\pi$ ratio as function of $p_{T}$ in $\sqrt{s_{NN}} = 200$ GeV central Au+Au , d+Au and near, all azimuth and away side from a high $p_{T}$ jet in 20 $\%$ central Au+Au events.}
\label{fig:moneyPlot}
\end{minipage}
\end{figure}
\section{Methodology and Results}
\label{4}
We obtained the per trigger jet - proton (pion) $\Delta\phi$ correlations. The proton (pion) $p_{T}$ range used is 2.0 - 2.8 GeV/c. A mix event technique is used prior to normalization to correct for detector acceptance inefficiencies. The triangular geometrical correlation in $\Delta\eta$ was corrected for. No efficiency correction has yet been applied to the correlation data. We expect the proton and pion efficiencies to cancel at this transverse momentum. A simple ratio of correlation shapes already shows a smaller p/$\pi$ ratio in the trigger jet axis azimuth direction and its recoiling side as seen in figure \ref{fig:simpleRatio}. We extract the near and away side particle yields from the $\Delta\phi$ correlation by fitting it with a model that includes contributions from near and away side jet Gaussians and a constant background. The fit model used is $f(\Delta\phi) = B+A_{near}exp(-0.5(\Delta\phi / \sigma_{near})^{2})+A_{away}exp(-0.5((\Delta\phi - \pi) / \sigma_{away})^{2})$. An attempt to include an azimuth quadrupole component in the model did not improve the fit and gave a proton $\nu_{2}$ consistent with zero and a pion $\nu_{2}$  = 0.031 $\pm$ 0.015 (assuming that $\nu_{2}^{hadrons}$ $\sim$ $\nu_{2}^{jet}$ above 2.0 GeV/c). The small values of the quadrupoles might not be surprising given the high $p_{T}^{jet}$ of our trigger. The azimuth quadrupole was considered too small to affect the near and away side jet yields and was dropped from the fit model.
\\
\indent The combinatorial uncorrelated background B can be subtracted and the near and away side jet yield for protons and/or pions can be obtained by integrating the corresponding Gaussians over the entire azimuth region. The p/$\pi$ ratio obtained by integrating the complete fit function over the  $\Delta\phi$ range is plotted on figure \ref{fig:moneyPlot} as a black square. Other STAR measurements are shown for comparison \cite{PhysRevLett.97.152301} .This ratio is consistent with the one obtained from inclusive Au+Au 0-12 $\%$ most central collisions. The near side ratio is suppressed (shown as a red star) and similar to the ratio in d+Au minimun bias collisions.  The away side ratio (shown as a blue triangle) is higher than the near side ratio but lower than the inclusive central Au+Au one. A comparison with jet-triggered peripheral Au+Au collisions and jet-triggered p+p collisions is in progress. 

\section{Conclusion}
\label{5}
The p/$\pi$ ratios at intermediate $p_{T}$ correlated with a jet trigger ($p_{T} \sim$ 8- 20 GeV/c) in the 20 $\%$ most central 200 GeV Au+Au collisions are analyzed. The near side p/$\pi$ ratio is significantly lower than the bulk ratio reflecting a near side jet property. The away side p/$\pi$ ratio is larger than the near side and seems smaller than the bulk ratio. This result raises the question whether this is an away side - medium interaction effect. A definite conclusion awaits results of peripheral Au+Au collisions and p+p collisions and a full assessment of systematic uncertainties, both in progress.
\bibliographystyle{elsarticle-num}
\bibliography{proceedingsDavilaHP2012}
\end{document}